\documentclass[12pt,preprint]{aastex}

\def\kms{\ifmmode{\rm km\th s^{-1}}\else km\th s$^{-1}$\fi}
\def\th{\thinspace}

\shortauthors{Boden et al.}
\shorttitle{DQ~Tau}

\begin{document}

\title{Interferometric Evidence for Resolved Warm Dust in the DQ Tau System}

%%%\email{**Submission 4 -- 12 Mar 2009**}

\author{Andrew F.\ Boden\altaffilmark{1,2},
        Rachel L.\ Akeson\altaffilmark{2},
        Anneila I.\ Sargent\altaffilmark{1},
        John M.\ Carpenter\altaffilmark{1}, \\
	David R.~Ciardi\altaffilmark{2},
	Jeffrey S.~Bary\altaffilmark{3,4},
	Michael F.~Skrutskie\altaffilmark{4},
}

%%%	Rafael Millan-Gabet\altaffilmark{1},

\altaffiltext{1}{Division of Physics, Math, and Astronomy, California
Institute of Technology, MS 105-24, Pasadena, CA 91125}

\altaffiltext{2}{NASA Exoplanet Science Institute, California
Institute of Technology, 770 South Wilson Ave., Pasadena CA 91125}

\altaffiltext{3}{Department of Physics \& Astronomy, Colgate University
Hamilton NY 13346}

\altaffiltext{4}{Department of Astronomy, University of Virginia,
Charlottesville, VA 22902}

\email{bode@astro.caltech.edu}

\begin{abstract}

We report on near-infrared (IR) interferometric observations of the
double-lined pre-main sequence (PMS) binary system DQ~Tau.  We model
these data with a visual orbit for DQ~Tau supported by the
spectroscopic orbit \& analysis of \citet{Mathieu1997}.  Further,
DQ~Tau exhibits significant near-IR excess; modeling our data requires
inclusion of near-IR light from an 'excess' source.  Remarkably the
excess source is resolved in our data, similar in scale to the binary
itself ($\sim$ 0.2 AU at apastron), rather than the larger
circumbinary disk ($\sim$ 0.4 AU radius).  Our observations support
the \citet{Mathieu1997} and \citet{Carr2001} inference of significant
warm material near the DQ~Tau binary.
%%, and expected
%%dynamical clearing is not completely successful in dispersing this
%%material.

\end{abstract}


\section{Introduction}
\label{sec:introduction}

\objectname[DQ Tau]{DQ~Tau} (HBC~72) is among a small number of known
``classical'' T-Tauri (CTTS) spectroscopic binaries.  DQ~Tau's strong
H$\alpha$ emission and de facto T Tauri status was reported by
\citet{Joy1949}.  The system has a spectral energy distribution (SED)
typical of CTTS, with strong near and mid-infrared (IR) excess
\citep[][\S~\ref{sec:SED}]{Strom1989, Skrutskie1990, Mathieu1997}.
H$\alpha$ emission and continuum veiling indicate significant
accretion onto the stars \citep{Valenti1993,Hartigan1995,Basri1997}.
\citet[][herein M1997]{Mathieu1997} established DQ~Tau as an
eccentric, short-period (15.8 d), double-lined spectroscopic binary
composed of similar late K-stars; presumably the mid-IR excess is due
to emission from a circumbinary disk.  Further, M1997 and
\citet[][herein B1997]{Basri1997} identified photometric and
spectroscopic variability at the orbit period and phased near
periastron, interpreting these variations as enhanced accretion as the
stellar components encounter streams of infalling material from the
circumbinary disk.  Recently \citet{Salter2008} reported a mm flare in
DQ~Tau, but interpret the flare as interacting stellar magnetospheres
similar to that seen in V773~Tau~A \citep{Massi2008}.

Binary systems are expected to clear inner gaps in their circumbinary
disks (out to several times the binary semi-major axis) as material is
dynamically ejected by the components
\citep[][]{Artymowicz1994,Pichardo2005}.  \citet{Jensen1997} report
SED evidence for such inner-disk clearings in a set of T Tauri
binaries.  The SEDs in these systems show a deficit of near-IR
emission relative to mid-IR flux -- interpreted as a lack of the
warmer inner-disk material that has been dispersed by the stars'
orbital motion.  Conversely, the DQ~Tau SED (and a few others,
e.g.~AK~Sco \citep{Jensen1997}, and UZ~Tau~E \cite{Jensen2007})
exhibits no such near-IR deficit, leading M1997 to conclude ``there is
clearly warm material within the binary orbit.''  M1997 modeled the
DQ~Tau SED with a modest (5$\times$10$^{-10}$ M$_{\sun}$) amount of
warm (1000 K), optically-thin material in the DQ~Tau binary region.
\citet[][herein C2001]{Carr2001} supported this conclusion with
IR spectroscopic detection of warm (1200 K) CO gas emission from
DQ~Tau.

The presence of material inside the expected DQ~Tau dynamical gap is
unsurprising: the system shows strong accretion diagnostics.  But the
amount and morphology of this `inner' material bears on how such PMS
binary systems interact with and accrete material from their
circumbinary reservoir, and motivates study with the
highest-resolution techniques available.  Here we report on
observations of DQ~Tau with the Keck Interferometer
\citep[KI,][]{Colavita2003}.  These observations partially resolve the
DQ~Tau system, and allow us to model the system visual orbit based on
the spectroscopic orbit by M1997.  Further, we find that we must
account for a static `excess' flux which is compact, but partially
resolved in these data.  We interpret this excess source with possible
morphological models for the warm material postulated by M1997 and
C2001, and discuss implications on inner material and accretion in the
DQ~Tau system.

\section{DQ~Tau Spectral Energy Distribution}
\label{sec:SED}

\begin{figure}[ph]
\epsscale{0.85}
%%\plotone{figures/dqTau.sedPlots.ps}
\plotone{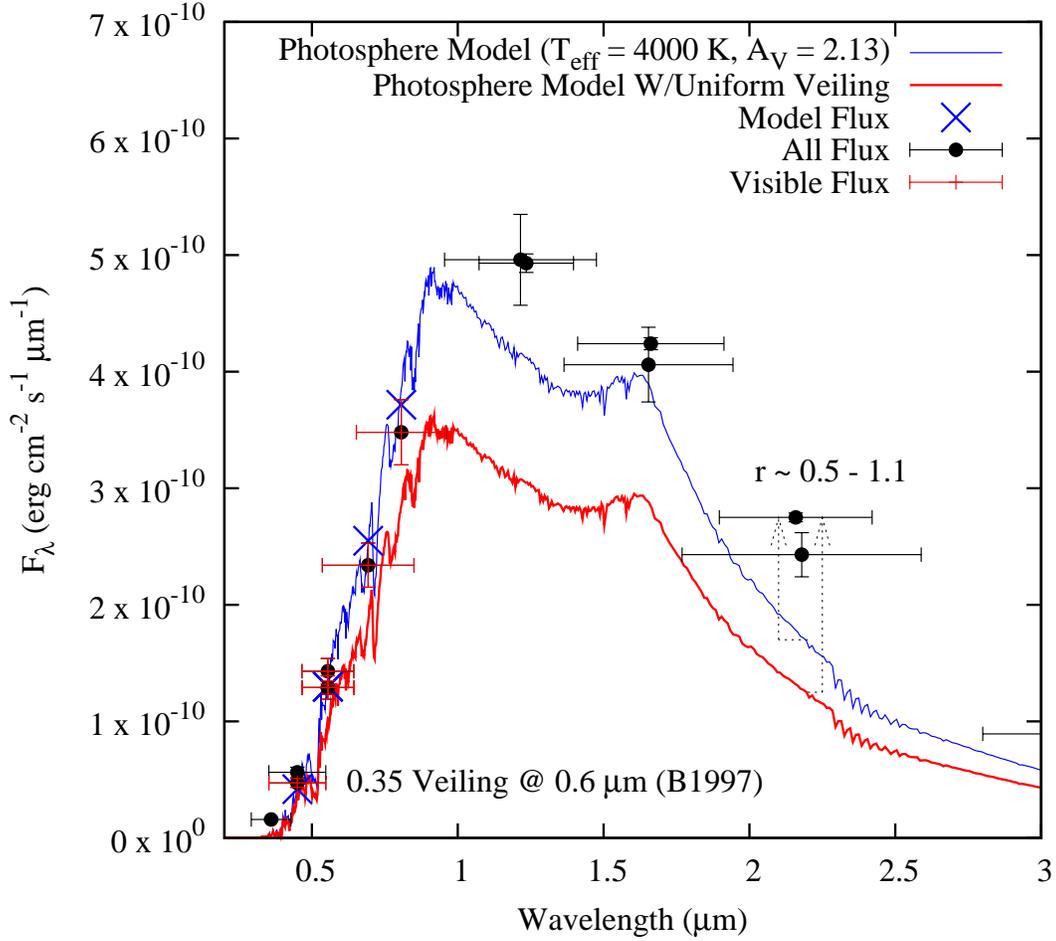}
\caption{DQ~Tau SED Model.  We model the DQ~Tau photospheres with a
  4000 K template (M1997) and significant extinction \cite[A$_V$ =
    2.13;][]{Strom1989}.  The red curve shows an alternative
  photospheric model assuming a veiling fraction of 0.35 $\pm$ 0.15
  (B1997).  The fractional $K$-excess over the photosphere (r) ranges
  between 0.5 and 1.1 depending on which photosphere model is adopted.
\label{fig:SEDplot}}
\end{figure}

The DQ~Tau SED has been studied by \citet{Strom1989},
\citet{Skrutskie1990}, \cite{Hartigan1995}, and M1997; as it bears on
our KI data analysis we summarize the SED here.
Figure~\ref{fig:SEDplot} shows a visible and near-IR SED model for the
DQ~Tau stellar photospheres derived using photometry from
\citet{Strom1989} and 2MASS \citep{Skrutskie2006}.  Adopting published
extinction estimates \citep[A$_V$ = 2.13;][]{Strom1989} and
photospheric parameters ($T_{\rm eff}$ = 4000 K, log g = 4.0, solar
abundance; M1997), we can model photometry between 0.45 and 1 $\mu$m
(excluding blue accretion flux and IR excess) with a single
photosphere (the stellar components of DQ~Tau are nearly identical;
M1997).  We find the net photosphere model shown in
Figure~\ref{fig:SEDplot}, corresponding to a total stellar luminosity
of 0.88 L$_\sun$ for D = 140 pc, in good agreement with previous
estimates \citep[e.g.][M1997]{Strom1989}.  However, these estimates
ignore the significant (0.35 $\pm$ 0.15 at 0.6 $\mu$m) veiling
reported by B1997.  An alternative model assuming uniform veiling
between 0.45 and 1 $\mu$m is also given in Figure~\ref{fig:SEDplot},
corresponding to total stellar luminosity of 0.65 L$_\sun$.  The total
luminosity of the DQ~Tau photospheres is likely between these two
extremes.

DQ~Tau's SED exhibits strong IR excess beyond 1 $\mu$m.  M1997 modeled
this excess with a power-law, and particularly argued for the presence
of a cooler circumbinary disk and additional warm material within the
binary orbit based on substantial excess over the stellar emission
between 1 -- 5 $\mu$m.  Most significant for our purposes is the
excess in $K$-band (2.2 $\mu$m) -- highlighted in
Fig~\ref{fig:SEDplot}.  Based on the photospheric models from above we
estimate the fractional excess $r$ at $K$ over the stellar
contribution to be in the range of 0.5 -- 1.1 (modulo intrinsic
$K$-variability of the system; we return to this question in
\S~\ref{sec:offset}).  The bottom of this range agrees with similar
estimates from \cite{Strom1989} and \citet{Skrutskie1990}, while
higher values result from encorporating the veiling estimate of B1997.
This SED model makes it clear our KI observations will contain flux
from both the DQ~Tau binary components and additional material, but
the exact value of the excess $K$-emission over stellar is uncertain
to a factor of two.

\section{Observations and Orbital Modeling}
\label{sec:observations}

\begin{figure}[ph]
\epsscale{0.9}
%%\plotone{figures/fit3.v2.strip.ps}
\plotone{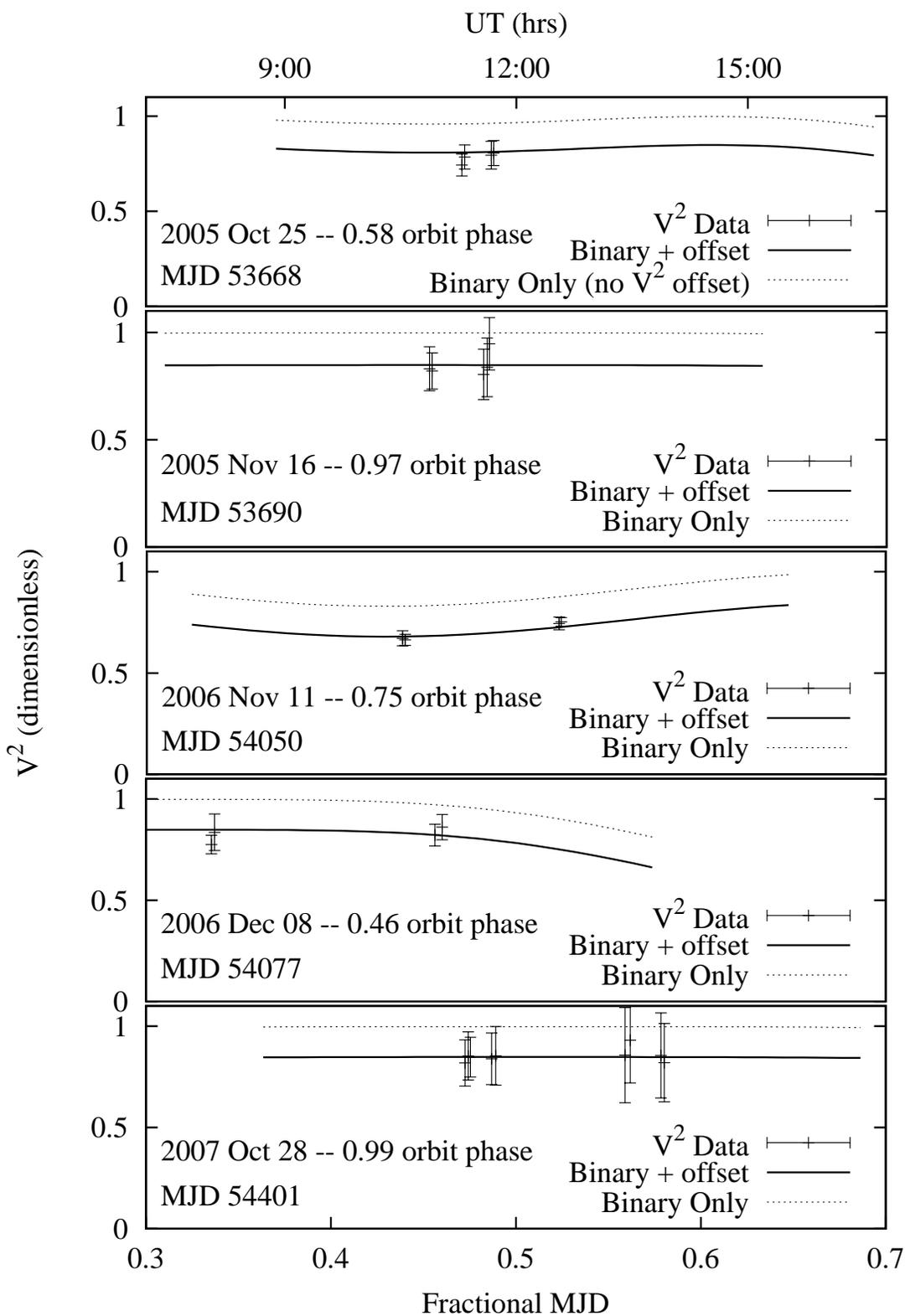}
\caption{KI Data/Model Comparisons for DQ~Tau V$^2$ Observations.
Here we give comparisons for our five epochs of KI V$^2$ observations
and our best-fit orbital model with and without the estimated
visibility offset.  In each frame both KI data and model are shown.
\label{fig:v2Data}}
\end{figure}

\paragraph{KI Observations}
The KI observable used for these measurements is the fringe contrast
or {\em visibility} (specifically, power-normalized visibility modulus
squared, V$^2$) of an observed brightness distribution on the sky.  KI
observed DQ~Tau in $K$-band on five nights between 2005 Oct 25 and
2007 Oct 28, a data set spanning roughly two years and 49 orbital
periods.  DQ~Tau and calibration objects were typically observed
multiple times during each of these nights, and each observation
(scan) was approximately 130 sec long.  As in previous publications,
our KI V$^2$ calibration follows standard procedures described in
\citet{Colavita2003}.  For this analysis we use \objectname[HD
  27282]{HD~27282} (G8 V) as our calibration object, resulting in 26
calibrated visibility scans on DQ~Tau over five epochs.  The V$^2$
observations are depicted in Figure~\ref{fig:v2Data}, along with our
best-fit orbit model (discussed below).  Orbital analysis methods for
such V$^2$ observations are discussed in \citet{Boden2000} and not
repeated here.

Figure~\ref{fig:v2Data} shows the DQ~Tau calibrated visibilities are
significantly less than one in all five epochs -- indicating resolved
structure in our observations.  Further, the five epochs show
variability with time and hour angle -- indicating the source
appearance changes with time (e.g. with phase of the DQ Tau binary
orbit), and is non-axisymmetric (both as expected for a resolved
binary system).

\paragraph{Orbit Model}

As in previous analyses \citep[e.g.][]{Boden2000,Boden2005,Boden2007}
we model the KI visibilities with a binary source.  The DQ~Tau
physical parameters from M1997 and putative 140 pc distance imply an
apparent semi-major axis on the order of 1 mas.  This apparent
separation is marginally resolved in our data (projected fringe
spacing of 5.1 mas), and does not allow an independent solution for
the binary visual orbit.  Therefore we have constrained our orbital
modeling with parameters from M1997 \citep[with period slightly
  revised by][]{Huerta2005}.  Effectively we solve only for $\Omega$
and the sense of rotation on the sky (i.e.~whether $i$ is greater or
less than 90$^\circ$).  Further, from \S~\ref{sec:SED} it is
necessary to account for the visibility contribution of DQ~Tau's $K$
excess (even if that flux is incoherent on angular scales measured in
these data).  We find it sufficient to model the visibility due to the
excess flux with a single, static visibility offset parameter
$V^{2}_{\rm offset}$ (discussed in \S\ref{sec:offset}).

\begin{figure}[t]
\epsscale{1.0}
%%\plotone{figures/dqTau.orbitTrace.eps}
\plotone{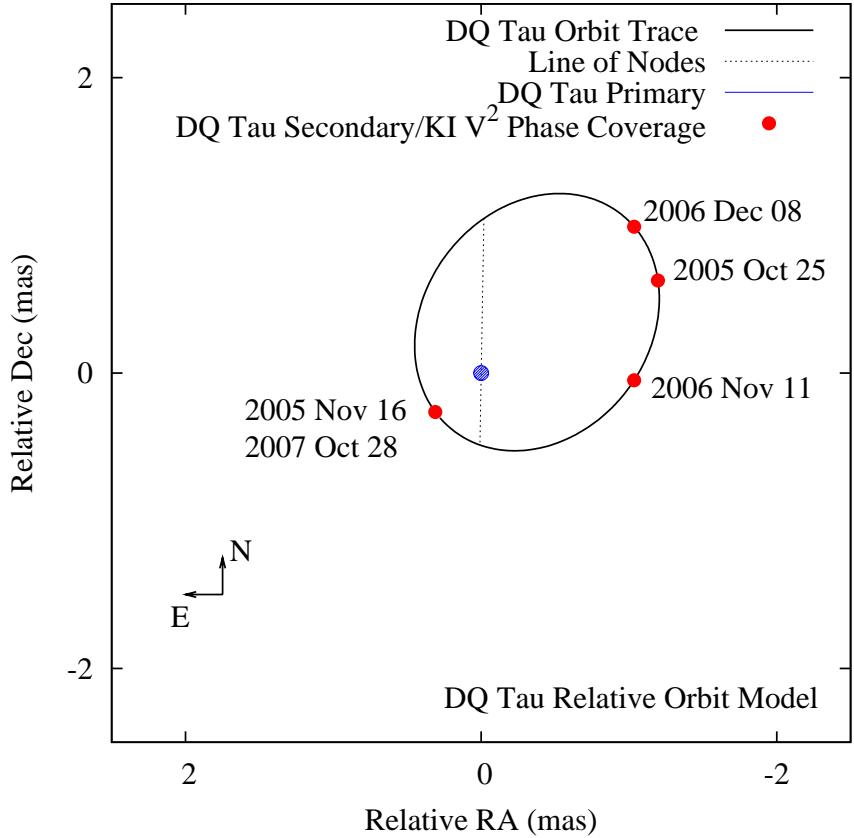}
\caption{Relative Visual Orbit of DQ~Tau.  The relative visual orbit
of the DQ~Tau system is given, with the primary rendered at the
origin, and the secondary position shown at the five epochs/orbit
phases of our observations (two of our epochs are at periastron).
\label{fig:orbitPlot}}
\end{figure}

Figure~\ref{fig:orbitPlot} depicts our relative visual orbit model,
with the primary rendered at the origin, and the secondary rendered at
the five orbit phases of our KI data.  Figure~\ref{fig:v2Data} shows
the comparison of the KI data and predictions; the model clearly
matches the data well within the estimated errors.
Table~\ref{tab:orbit} summarizes our DQ~Tau orbit model as adopted
from M1997 and derived here.  Our data show that the orbit motion is
clockwise on the sky (retrograde); we adopt an inclination value
(157$^\circ$) constrained by the M1997 $\sin i$ estimate and
reflecting retrograde orbit motion.

\begin{deluxetable}{lcc}
%%%\tabletypesize{\footnotesize}
\tablecolumns{3}
\tablewidth{0pc}
%%\rotate

\tablecaption{Orbital Parameters for DQ~Tau.
\label{tab:orbit}
}

\tablehead{
\colhead{Orbital}        & \colhead{M1997}      & This Work \\
\colhead{Parameter}      &                      & 
}
\startdata
Period (d)               & 15.8016              &                          \\
T$_{0}$ (MJD)            & 49582.04             &                          \\
$e$                      & 0.556                &                          \\
K$_{Aa}$ (km s$^{-1}$)   & 21.6                 &                          \\
K$_{Ab}$ (km s$^{-1}$)   & 22.4                 &                          \\
$\gamma$ (km s$^{-1}$)   & 22.4                 &                          \\
$\omega_{A}$ (deg)       & 228                  &                          \\
$\Omega$ (deg)           &                      & 179 $\pm$ 10             \\
$i$ (deg)                &  {\em 23}            & {\em 157}                \\
$a$ (mas)                &                      & {\em 0.96}               \\
$\Delta$ $K$ (mag)       &                      & {\em 0}                  \\
V$^2_{\rm offset}$       &                      & 0.15 $\pm$ 0.03          \\
\hline
\enddata

\tablecomments{Our orbit model is constrained to parameters estimated
  by M1997 except for the period \citep{Huerta2005}, $\Omega$ and
  V$^2_{\rm offset}$.  Note that the inclination $i$ is constrained to
  the M1997 value, except that our data \& modeling indicate the
  motion of the binary is clockwise (retrograde) on the plane of the
  sky.}

\end{deluxetable}

\section{Interpreting the Visibility Offset}
\label{sec:offset}

$V^{2}_{\rm offset}$ accounts for contributions from DQ~Tau's $K$
excess in our modeling.  We have taken this visibility offset as
static, that is invariant in time and with baseline projection angle;
this construct warrants some consideration.  First we note that M1997
and B1997 %%\citet{Basri1997}
reported photometric variability in DQ~Tau, but
this was strongest in blue (U \& B) colors and attributed to
variability in accretion luminosity; as we will argue below
spectrophotometric monitoring suggests changes in the system $K$ flux
are modest.  Second, the assumption of an offset that is independent
of baseline projection de facto assumes that the $K$ excess centroid
is centered (co-axial) on the binary center of mass/light (for an
equal-mass binary), and is itself axially symmetric.
\citet{Artymowicz1996} have modeled accretion in systems such as
DQ~Tau, and their modeling clearly shows time-variable and
non-axisymmetric structures.  These time-variable features call into
question our static visibility offset construct, but the degree of
visibility variability will depend on the relative intensity of
symmetric and non-symmetric components of the $K$ flux in the DQ~Tau
excess, and the spatial frequency coverage in our interferometric
data.  In our last epoch (2007 Oct 28; Figure~\ref{fig:v2Data} bottom)
we specifically made the observation at periastron, and pushed the KI
instrument to its maximum range in hour angle (roughly four hours) to
test this static offset/axisymmetic emission assumption to the
greatest extent possible.  Data from this last epoch show no
significant signs of variation with hour angle, supporting the
axisymmetric emission construct.  Given these considerations we settle
on a static visibility offset to account for the $K$ excess because
nothing more complicated is justified by our data: the binary plus
static offset construct adequately model our KI data over their range
of spatial frequency.

The (monochromatic) V$^2$ of a two-component composite scene is given
by:
\begin{displaymath}
V^{2}_{\rm composite}
 = \frac{V^{2}_{1} + r^2 V^{2}_{2} + 2 r V_{1} V_{2} \cos{\phi}}
        {(1 + r)^2} 
 \rightarrow \frac{(V_{1} + r V_{2})^2}{(1 + r)^2} 
\end{displaymath}
with $V_{1}$ and $V_{2}$ the visibilities of two components, $r$ the
flux ratio (2 to 1), and $\phi$ the phase difference between the two
fringes (e.g. $\phi$ = 2 $\pi/\lambda$ ${\bf B} \cdot {\bf s}$ for a
typical binary source); the second form assumes this phase offset is
zero or the two sources are co-axial.  Identifying $V_{1}$ and $V_{2}$
with the DQ~Tau binary and the $K$-excess respectively, and evaluating
the expression at a convenient binary phase (i.e. periastron, where
$V_{\rm binary} \approx 1$) leaves:
\begin{displaymath}
V^{2}_{\rm composite} ({\rm periastron})
\approx \frac{(1 + r V_{\rm excess})^2}{(1 + r)^2} 
\approx 1 - V^{2}_{\rm offset}
\end{displaymath}
This allows us to estimate the net visibility of the DQ~Tau $K$-excess
in terms of V$^{2}_{\rm offset}$ (Table~\ref{tab:orbit}) as:
\begin{equation}
V_{\rm excess} = \frac{1}{r}\left((1 - V^{2}_{\rm offset})^{(1/2)} (1 + r) - 1\right)
\label{eq:excessVisibility}
\end{equation}
Evaluating Eq.~\ref{eq:excessVisibility} with $r$ in the range of 0.5
-- 1.1 (\S~\ref{sec:SED}) and $V^{2}_{\rm offset}$ = 0.15
(Table~\ref{tab:orbit}) yields $V_{\rm excess}$ = 0.77 -- 0.85.

M1997 documents significant optical variability in DQ~Tau, and
attributes the variability to enhanced accretion near periastron.
Pertinent to this discussion is the degree of $K$ variability in the
system.  DQ~Tau has been monitored with the CorMASS instrument
\citep{Wilson2001} as part of a T~Tauri accretion variability program
discussed in \citet{Bary2008}.  Analysis of these data suggests that
$K$ emission during enhanced accretion is approximately 20\% brighter
than during a quiescent phase.  This variation is small compared to
the factor of two uncertainty in $r$ from the DQ~Tau photospheric
uncertainty.

Remarkably this range for $V_{\rm excess}$ indicates the KI data
resolve, but do not over-resolve, the $K$ excess, and this result is
robust even incorporating the large uncertainty in $r$.  For instance,
if the dominant $K$-excess came from a warm inner-edge of the
circumbinary disk \citep[nominally at a barycentric distance of
  $\approx$ 0.4 AU;][]{Artymowicz1994,Pichardo2005}, the apparent
inner-edge diameter would be $\sim$ 5.7 mas.  The $V$ of such a ring
morphology would be $\sim$ 0.1 -- 0.2.  Apparently the $K$ excess is
dominated by emission from a region significantly more compact than
the dynamically-allowed inner edge of the circumbinary disk.  Our
visibility measurement supports the M1997 and C2001 conclusion that
there is warm material in the vicinity of the DQ~Tau binary, and
emission from that material dominates the system's $K$ excess.

If we neglect $K$-emission from the circumbinary disk entirely (the
equilibrium temperature of material at 0.4 AU is on the order of
475~K), we can estimate the angular scale of emission from the warm
material.  Table~\ref{tab:simpleModels} summarizes a small set of
simple assumed emission morphologies for the $K$-excess: Gaussian
profile, thin ring, and uniform disk.  Under these assumptions the
characteristic apparent size required to match the $V_{\rm excess}$
estimate is given.  Of these three models we expect the Gaussian
profile most closely matches the warm material emission profile; we
note that this assumption results in a size scale (1.1 -- 1.4 mas;
0.15 -- 0.2 AU at 140 pc) that closely matched the apparent binary
separation at apastron ($\sim$ 1.5 mas).
%%, as
%%material near the binary barycenter will experience the most intense
%%radiation field, and thus be warmest, with temperatures (and
%%consequently emission) gradually falling off with distance from the
%%barycenter.
%%We note that this Gaussian-profile assumption results in a size scale
%%(1.4 mas; 0.2 AU at 140 pc) that closely matched the apparent binary
%%separation at apastron ($\sim$ 1.5 mas).
Incoherent scattered-light from the circumbinary disk \citep[proposed
  for CTTS by][]{Pinte2008} would make this warm material
characteristic size estimate smaller.

\begin{deluxetable}{lcc}
%%%\tabletypesize{\footnotesize}
\tablecolumns{3}
\tablewidth{0pc}
%%\rotate

\tablecaption{Characteristic Angular Scales for Simple Morphological Models of DQ~Tau $K$-band Excess.
\label{tab:simpleModels}
}

\tablehead{
\colhead{Model}          & Characteristic   & Note \\
                         &  Scale (mas)     & 
}
\startdata
Gaussian                 & 1.1 -- 1.4       & Gaussian FWHM            \\
Ring                     & 1.3 -- 1.7       & Ring Diameter            \\
Uniform Disk             & 1.9 -- 2.4       & UD Diameter              \\
\hline
\enddata

\tablecomments{Assuming $V_{\rm excess}$ = 0.85 -- 0.77}

\end{deluxetable}

\section{Discussion}

We have modeled our DQ~Tau observations based on orbital parameters
from M1997 and system SED.  Unsurprisingly, the significant $K$-excess
must be included in the V$^{2}$ modeling.  Remarkably these data
indicate the excess must come from a region smaller than the
circumbinary disk.
%%  This finding confirms the inference from M1997 and
%% C2001 that there is substantial warm material within the DQ~Tau orbit.
Further, our data suggest that this excess is distributed on the
physical scale of the binary orbit ($\sim$ 0.1 -- 0.2 AU) rather than
being either much smaller (e.g.~circumstellar disks) or much larger
(i.e.~the circumbinary disk) than the stellar separation.

Our data and modeling support the M1997 and C2001 inference that
DQ~Tau has significant warm material in the inner orbit region (in
addition to the substantial circumbinary disk).  Binary dynamics,
accretion, and wind/outflow processes work to dissipate this inner
material over a few binary orbital periods.  The static visibility
offset that adequately models our KI data over many DQ~Tau orbital
periods suggests the system is in quasi-equilibrium with material
inflow replenishing dissipated material in the binary region.

It is important to note that our present data on and modeling of this
remarkable system is relatively crude.  In particular the $\sim$ 5 mas
KI fringe spacing only partially resolves the $\sim$ 1 mas DQ~Tau
binary orbit.  Clearly limited spatial information leads to
limitations in our DQ~Tau modeling, and some care in interpreting our
conclusions is warranted.  Our construct that the $K$-excess
morphology is axisymmetric seems the most suspect; it runs counter to
existing modeling of accreting binary systems
\citep[e.g.][]{Artymowicz1996}, and photometric/accretion brightening
near periastron (M1997).  However nothing more sophisticated is
justified by our data -- in particular the KI data from our last epoch
(2007 Oct 28) test the axisymmetric modeling assumption to the
practical limits of KI capabilities (in broad-band data).
But there should be an asymmetric component to the near-IR flux at
some contrast level and spatial scale in the DQ~Tau system.  Going
forward it is important to probe the extent of any axisymmetry in the
near-IR excess at a greater diversity of spatial scales and
wavelengths to understand the distribution and flow of material in
DQ~Tau and similar systems.

\acknowledgements

Data presented herein were obtained at the W.M.~Keck Observatory.  We
gratefully acknowledge the support of personnel at the Jet Propulsion
Laboratory, W.M.~Keck Observatory, and the NASA Exoplanet Science
Center in KI observations of DQ~Tau.  We thank the anonymous referee
for his comments on including the veiling luminoisty in the SED
modeling for DQ~Tau.

\clearpage

\end{document}